\documentstyle[12pt]{article}
\def\zid{1\kern-0.36em\llap~1}

\newcommand{\beq}{\begin{equation}}
\newcommand{\ber}{\begin{eqnarray}}
\newcommand{\eeq}{\end{equation}}
\newcommand{\eer}{\end{eqnarray}}

\begin{document}

\begin{titlepage}
%\vbox {\vspace{0.1mm}} %Leaves space at top of 1st page.
\rightline{[SUNY BING 10/24/04] } \rightline{ hep-ph/0400000}
\vspace{2mm}
\begin{center}
{\bf \hspace{01.4 cm} ON AN INTENSITY-RATIO EQUIVALENCE FOR
\newline TWO TOP-QUARK DECAY COUPLINGS}\\
\vspace{2mm} Charles A. Nelson\footnote{Electronic address:
cnelson @ binghamton.edu  } \\ {\it Department of Physics, State
University of New York at Binghamton\\ Binghamton, N.Y.
13902}\\[2mm]
\end{center}

%\vspace{2mm}

\begin{abstract}

For the $t \rightarrow W^{+} b $ decay mode, an intensity-ratio
equivalence for two distinct Lorentz-invariant couplings is shown
to be a consequence of symmetries of tWb-transformations. Explicit
tWb-transformations, $A_{+}=M $ $A_{SM}$, $P $ $A_{SM}$, $B $
$A_{SM}$ connect the four standard model's (SM) helicity
amplitudes, $A_{SM}\left( \lambda _{W^{+} } ,\lambda _b \right)$,
and the amplitudes $A_{+}\left( \lambda _{W^{+} } ,\lambda _b
\right)$ in the case of an additional chiral-tensorial-coupling of
relative strength $\Lambda_{+} =E_W /2 \sim 53 GeV$. Such a
coupling will arise if there is a large $t_R \rightarrow b_L$
chiral weak-transition-moment.  Two commutator plus
anti-commutator symmetry algebras are generated from $ M, P, B$.
Using these transformations, the associated mass scales are
related to the SM's electroweak scale $v_{EW} \sim 246GeV$.

\end{abstract}

\end{titlepage}

\section{Introduction:}

In this paper, for the $t \rightarrow W^{+} b $ decay mode [1], an
intensity-ratio equivalence [2] for two distinct Lorentz-invariant
couplings is shown to be related to symmetries of
tWb-transformations, $A_{+}=M $ $A_{SM}$, $P$ $A_{SM}$, $B $
$A_{SM}$, where $ M, P, B$ are explicit $4$x$4$ matrices. These
tWb-transformations connect the standard model's helicity
amplitudes, $A_{SM}\left( \lambda _{W^{+} } ,\lambda _b \right)$,
and the amplitudes $A_{+}\left( \lambda _{W^{+} } ,\lambda _b
\right)$ in the case of an additional chiral-tensorial-coupling of
relative strength $\Lambda_{+} =E_W /2 \sim 53 GeV$.  Versus the
standard model's pure $(V-A)$ coupling, the additional $(f_M +
f_E)$ tensorial coupling can be physically interpreted as arising
due to a large $t_R \rightarrow b_L$ chiral weak-transition-moment
for the t-quark. $\Lambda_{+}$ is defined by (1) below and the
$(+)$ amplitudes' complete coupling is (2). $E_W$ is the energy of
the final W-boson in the decaying t-quark rest frame. The
subscripts $R$ and $L$ respectively denote right and left
chirality of the coupling, that is $( 1 \pm \gamma_5 ) $. $\lambda
_{W^{+} }$, $ \lambda _b $ are the helicities of the the emitted
W-boson and b-quark in the t-quark rest frame.   The Jacob-Wick
phase-convention [3] is used in specifying the phases of the
helicity amplitudes and so of these transformations.

As tests with respect to the most general Lorentz coupling, in
references [4,5], stage-two spin-correlation functions were
derived and studied as a framework for complete measurements of
the helicity parameters for $t \rightarrow W^{+} b $ decay. Such
tests are possible at the Tevatron [1], at the LHC [6], and at a
NLC [7]. Due to rotational invariance, there are four independent
$A\left( \lambda _{W^{+} } ,\lambda _b \right)$ amplitudes for the
most general Lorentz coupling. In this paper, a subset of the most
general Lorentz coupling is considered in which the subscript
``$i$" identifies the amplitude's associated coupling: ``$i=$ SM"
for the pure $(V-A)$ coupling, ``$i=(f_M + f_E)$" for the pure
$t_R \rightarrow b_L$ tensorial coupling, and ``$i=(+)$" for
$(V-A) + (f_M + f_E)$ with a t-quark chiral weak-transition moment
of fixed relative strength $\Lambda_{+} =E_W /2$ versus $g_L$.
Explicit expressions for the helicity amplitudes associated with
each ``$i$" coupling are listed in Sec. 2.

The Lorentz coupling involving both the SM's $(V-A)$ coupling and
an additional $t_R \rightarrow b_L $ weak-moment coupling of
arbitrary relative strength $\Lambda_{+}$ is $ W_\mu ^{*} J_{\bar
b t}^\mu = W_\mu ^{*}\bar u_{b}\left( p\right) \Gamma ^\mu u_t
\left( k\right) $ where $k_t =q_W +p_b $, and
\begin{equation}
\frac{1}{2} \Gamma ^\mu =g_L\gamma ^\mu P_L + \frac{g_{f_M + f_E}
} {2\Lambda _{+} }\iota \sigma ^{\mu \nu } (k-p)_\nu P_R
\end{equation}
where $P_{L,R} = \frac{1}{2} ( 1 \mp \gamma_5 ) $.  In $g_L =
g_{f_M + f_E} = 1$ units, when $\Lambda_{+} = E_W/2$ which
corresponds to the (+) amplitudes, the complete $t \rightarrow b$
coupling is very simple
\begin{equation}
\gamma ^\mu P_L + \iota \sigma ^{\mu \nu } v_\nu P_R
 =P_R \left( \gamma
^\mu + \iota \sigma ^{\mu \nu } v_\nu \right)
\end{equation}
where $v_{\nu}$ is the W-boson's relativistic four-velocity.

We denote by $\Gamma$ the partial-width for the $t \rightarrow
W^{+} b $ decay channel and by $\Gamma_{L,T}$ the partial-width's
for the sub-channels in which the $W^{+}$ is respectively
longitudinally, transversely polarized; $\Gamma = \Gamma_L
+\Gamma_T $.  Similarly, ${\Gamma_L}|_{\lambda_b = - \frac{1}{2}}
$ denotes the partial-width for the W-longitudinal sub-channel
with b-quark helicity $\lambda_b = - \frac{1}{2}$, etc.

The intensity-ratio equivalence statement is: ``As consequence of
Lorentz-invariance,  for the $t \rightarrow W^{+} b $ decay
channel each of the four ratios ${\Gamma_L}|_{\lambda_b = -
\frac{1}{2}} / {\Gamma}$, ${\Gamma_T}|_{\lambda_b = - \frac{1}{2}}
/ {\Gamma}$, ${\Gamma_L}|_{\lambda_b =  \frac{1}{2}} / {\Gamma}$,
${\Gamma_T}|_{\lambda_b =  \frac{1}{2}} / {\Gamma}$, is identical
for the pure $(V-A)$ coupling and for the $(V-A) + (f_M + f_E)$
coupling with $\Lambda_{+} =E_W /2$, and their respective
partial-widths are related by $\Gamma_{+} = v^2 \Gamma_{SM}$." $v
\simeq 0.65$ is magnitude of the the velocity of the W-boson in
the t-quark rest frame. The two couplings are obviously physically
distinct because their associated amplitudes have different
relative phases, see Table 1. This equivalence does not require
specific values of the mass ratios $y \equiv m_W/m_t$, and $x
\equiv m_b/m_t$, but $\Lambda_{+} =E_W /2$ does determine the
relative strength of the chiral weak-transition moment for the
t-quark versus $g_L$, as in (2) versus (1).

In the  $t$ rest frame, the helicity-amplitude matrix element for
$t \rightarrow W^{+} b$ is \newline $ \langle \theta _1^t ,\phi
_1^t ,\lambda _{W^{+} } ,\lambda _b |\frac 12,\lambda _1\rangle =$
 $ D_{\lambda _1,\mu }^{(1/2)*}(\phi _1^t ,\theta _1^t ,0)A_{i}
\left( \lambda _{W^{+} } ,\lambda _b \right) $ where $\mu =\lambda
_{W^{+} } -\lambda _b $ in terms of the $W^+$ and $b$-quark
helicities.  The asterisk denotes complex conjugation, the final
$W^{+}$ momentum is in the $\theta _1^t ,\phi _1^t$ direction, and
$\lambda_1$ gives the $t$-quark's spin component quantized along
the $z$ axis. $\lambda_1$ is also the helicity of the $t$-quark if
one has boosted, along the ``$-z$" direction, back to the $t$ rest
frame from the $(t \bar{t})_{cm}$ frame.  It is this boost which
defines the $z$ axis in the $t$-quark rest frame for angular
analysis [4].

The three tWb-transformations, $A_{+}=M $ $A_{SM}$, $P $ $A_{SM}$,
$B $ $A_{SM}$, are related to this equivalence statement.  As
explained in Sec. 2, the $M$ transformation implies the above
equivalence statement, but postulating M also implies the sign and
ratio differences of the (ii) and (iii) type amplitude
ratio-relations which distinguish the (SM) and (+) couplings. The
$P$ and $B$ transformations more completely exhibit the underlying
symmetries relating these two Lorentz-invariant couplings. In
particular, these three $4$x$4$ matrices lead to two ``commutator
plus anti-commutator" symmetry algebras, and together can be used
to relate the values of $\Lambda_{+}$, $m_t$, and $m_b$ to the
SM's electroweak scale $v_{EW}=\sqrt{-\mu^2 / \vert \lambda \vert}
= \sqrt{2} \langle 0|\phi |0\rangle \sim 246GeV$ where $\phi$ is
the Higgs field.

In Sec. 2, it is shown how these three tWb-transformations
successively arise from consideration of different types of
``helicity amplitude relations" for $t \rightarrow W^{+} b $
decay: The type (i) are ratio-relations which hold separately for
the two cases, ``$i=(SM)$, $(+)$". The type (ii) are
ratio-relations which relate the amplitudes in the two cases. From
the type (iii) ratio-relations which also relate the two cases,
the tWb-transformation $A_{+}=M$ $A_{SM}$ is introduced, where
$M=v$ $diag(1,-1,-1,1)$ characterizes the mass scale $\Lambda_{+}
= E_W/2 $. The amplitude condition (iv)
\begin{equation}
A_{+} (0,-1/2) = a A_{SM} (-1,-1/2),
 \end{equation}
with $a= 1 + O(v \neq y \sqrt{2}, x)$, determines the scale of the
tWb-transformation matrix $P$ and determines the value of the mass
ratio $y \equiv m_W/m_t$. $O(v \neq y \sqrt{2}, x)$ denotes small
corrections, see below. The amplitude condition (v)
\begin{equation}
A_{+} (0,-1/2) = - b A_{SM} (1-1/2),
\end{equation}
with $ b = v^{-8} $, determines the scale of $B$ and determines
the value of $ x= m_b/m_t$.  In Sec. 3, the two symmetry algebras
are obtained which involve the $M$, $P$, and $B$ transformation
matrices. Sec. 4 contains a discussion of these results and their
implications assuming that the observed $t \rightarrow W^{+} b $
decay mode will be found empirically to be well-described by (2).

\section{Helicity amplitude relations:}

In the Jacob-Wick phase convention, the helicity amplitudes for
the most general Lorentz coupling are given in [4]. In $g_L =
g_{f_M + f_E} = 1$ units and suppressing a common overall factor
of $\sqrt{m_t \left( E_b +q_W \right) }$, for only the $(V-A)$
coupling the associated helicity amplitudes are:
\begin{eqnarray*}
 A_{SM} \left(
0,-\frac 12\right) & = & \frac{1 }{y } \; \frac{E_W +q_W }{m_t }
\\
 A_{SM} \left(
-1,-\frac 12\right) & = & \sqrt{2}  \\
 A_{SM} \left( 0,\frac 12\right)
& = & -  \frac{1 }{y }  \frac{E_W -q_W }{m_t } \left( \frac
{m_b}{m_t-E_W +  q_W}  \right) \\
 A_{SM} \left( 1,\frac 12\right) & = & -
\sqrt{2} \left( \frac {m_b}{m_t-E_W +  q_W}  \right)
\end{eqnarray*}
For only the $(f_M + f_E)$ coupling, i.e. only the additional $t_R
\rightarrow b_L $ tensorial coupling:
\begin{eqnarray*}
 A_{f_M + f_E} \left(
0,-\frac 12\right) & = &  - ( \frac{m_t }{2\Lambda_+ }) \;  y  \\
 A_{f_M + f_E} \left(
-1,-\frac 12\right) & = & - ( \frac{m_t }{2\Lambda_+ }) \sqrt{2}
\; \frac{E_W +q_W }{m_t } \\
 A_{f_M + f_E} \left( 0,\frac 12\right)
& = & ( \frac{m_t }{2\Lambda_+ }) y \left( \frac {m_b}{m_t-E_W +
q_W}  \right) \\
 A_{f_M + f_E} \left( 1,\frac 12\right) & = & ( \frac{m_t }{2\Lambda_+ })
\sqrt{2} \;  \frac{E_W - q_W }{m_t } \left( \frac {m_b}{m_t-E_W +
 q_W}  \right)
\end{eqnarray*}
From these, the amplitudes for the $(V-A) + (f_M + f_E)$ coupling
of (1) are obtained by \newline $A_{+} (\lambda_W, \lambda_b) =
A_{SM} (\lambda_W, \lambda_b) + A_{f_M + f_E } (\lambda_W,
\lambda_b)$. For $\Lambda_{+} = E_W /2$, the $A_{+} (\lambda_W,
\lambda_b)$ amplitudes corresponding to the complete $t
\rightarrow b$ coupling (2) are
\begin{eqnarray*}
 A_{+} \left(
0,-\frac 12\right) & = & \frac{1 }{y } \; (q/E_W)  \; \frac{E_W
+q_W }{m_t }
\\
 A_{+} \left(
-1,-\frac 12\right) & = & - \sqrt{2} \; (q/E_W) \\
 A_{+} \left( 0,\frac 12\right)
& = & \frac{1 }{y } \; (q/E_W) \;  \frac{E_W -q_W }{m_t } \left(
\frac {m_b}{m_t-E_W +  q_W}  \right) \\
 A_{+} \left( 1,\frac 12\right) & = & -
\sqrt{2} \; (q/E_W) \; \left( \frac {m_b}{m_t-E_W +  q_W}  \right)
\end{eqnarray*}
For each of the three ``$i$" couplings, a direct derivation from
(1) shows how the different factors arise in the amplitudes [8].

We now analyze the different types of helicity amplitude relations
involving both the SM's amplitudes and those in the case of the
$(V-A) + (f_M + f_E)$ coupling: The first type of ratio-relations
holds separately for $i=(SM)$, $(+)$ and for all $y = \frac {m_W}
{m_t} , x = \frac {m_b} {m_t} , \Lambda_{+}$ values, (i):
\begin{equation}
\frac{A_{i} (0,1/2) } { A_{i} (-1,-1/2) } = \frac{1}{2}
\frac{A_{i} (1,1/2) } { A_{i} (0,-1/2) }
\end{equation}

The second type of ratio-relations relates the amplitudes in the
two cases and also holds for all $y, x, \Lambda_{+}$ values. The
first two relations have numerators with opposite signs and
denominators with opposite signs, c.f. Table 1; (ii): Two
sign-flip relations
\begin{equation}
\frac{A_{+} (0,1/2) } { A_{+} (-1,-1/2) } = \frac{A_{SM} (0,1/2) }
{ A_{SM} (-1,-1/2) }
\end{equation}
\begin{equation}
\frac{A_{+} (0,1/2) } { A_{+} (-1,-1/2) } = \frac{1}{2}
\frac{A_{SM} (1,1/2) } { A_{SM} (0,-1/2) }
\end{equation}
and two non-sign-flip relations
\begin{equation}
\frac{A_{+} (1,1/2) } { A_{+} (0,-1/2) } = \frac{A_{SM} (1,1/2) }
{ A_{SM} (0,-1/2) }
\end{equation}
\begin{equation}
\frac{A_{+} (1,1/2) } { A_{+} (0,-1/2) } = 2 \frac{A_{SM} (0,1/2)
} { A_{SM} (-1,-1/2) }
\end{equation}
(7, 9), which are not in [2], are essential for obtaining the $P$
and $B$ tWb-transformations and thereby the symmetry algebras of
Sec. 3 below.

The third type of ratio-relations, holding for all $y, x$ values,
follows by determining the effective mass scale, $\Lambda_{+}$, so
that there is an exact equality for the ratio of left-handed
amplitudes (iii):
\begin{equation}
\frac{A_{+} (0,-1/2) } { A_{+} (-1,-1/2) } = -
 \frac{A_{SM} (0,-1/2) } { A_{SM} (-1,-1/2) },
\end{equation}
Equivalently, $ \Lambda_{+} = \frac{m_t } {4 } [ 1 + (m_W / m_t)^2
- (m_b / m_t)^2] = E_W/2$ follows from each of:
\begin{equation}
\frac{A_{+} (0,-1/2) } { A_{+} (-1,-1/2) } = -
 \frac{1}{2} \frac{A_{SM} (1,1/2) } { A_{SM} (0,1/2) },
\end{equation}
\begin{equation}
\frac{A_{+} (0,1/2) } { A_{+} (1,1/2) } = - \frac{A_{SM} (0,1/2) }
{ A_{SM} (1,1/2) },
\end{equation}
\begin{equation}
\frac{A_{+} (0,1/2) } { A_{+} (1,1/2) } = - \frac{1}{2}
\frac{A_{SM} (-1,-1/2) } { A_{SM} (0,-1/2) },
\end{equation}
From the amplitude expressions given above, the value of this
scale $\Lambda_{+}$ can be characterized by postulating the
existence of a tWb-transformation $A_{+}=M$ $A_{SM}$ where $M=v$
$diag(1,-1,-1,1)$, with $
A_{SM}=[A_{SM}(0,-1/2),A_{SM}(-1,-1/2),A_{SM}(0,1/2),A_{SM}(1,1/2)]$
and analogously for $A_{+}$.

Assuming (iii), the fourth type of relation is the equality (iv):
\begin{equation}
A_{+} (0,-1/2) = a A_{SM} (-1,-1/2),
\end{equation}
where $a= 1 + O(v \neq y \sqrt{2}, x)$.

This is equivalent to the W-boson velocity formula $  v = a  y
\sqrt{2} \left( \frac{1} {1- ( E_b - q_W )/m_t} \right)
 = a y \sqrt{2} $ for $ m_b = 0 $.
In [2], for $a=1$ it was shown that (iv) leads to a cubic equation
with the solution $y=\frac {m_W} {m_t} = 0.46006$ ($ x=0$). The
present empirical value is $ y = 0.461 \pm 0.014$, where the error
is dominated by the $3 \% $ precision of $m_t$. In [2], for $a=1$
it was also shown that (iv) leads to $\sqrt{2}=v\gamma
(1+v)=v\sqrt{\frac{1+v}{1-v}}$ so $v=0.6506\ldots$ without input
of a specific value for $m_b$. However, by Lorentz invariance $v$
must depend on $m_b$. Accepting (iii), we interpret this lack of
dependence on $m_b$ to mean that $a \neq 1$ and in the Appendix
obtain the form of the $O(v \neq y \sqrt{2}, x)$ corrections in
$a$ as required by Lorentz invariance.  The small correction $O(v
\neq y \sqrt{2}, x)$ depends on both $x \equiv m_b/m_t$ and the
difference $v-y \sqrt{2}$.

Equivalently, by use of (i)-(iii) relations, (14) can be expressed
postulating the existence of a second tWb-transformation $A_{+}=P
$ $A_{SM}$ where
\begin{equation}
P\equiv v
\left[
\begin{array}{cccc}
0 & a/v & 0 & 0 \\ -v/a & 0 & 0 & 0 \\ 0 & 0 & 0 & -v/2a \\ 0 & 0
& 2a/v & 0
\end{array}
\right]
\end{equation}
The value of the parameter $a$ of (iv) is not fixed by (15).

The above two tWb-transformations do not relate the $\lambda_b= -
\frac{1}{2}$ amplitudes with the $\lambda_b= \frac{1}{2}$
amplitudes. From (i) thru (iv), in terms of a parameter $b$, the
equality (v):
\begin{equation}
A_{+} (0,-1/2) = - b A_{SM} (1,1/2),
\end{equation}
is equivalent to $A_{+}=B $ $A_{SM}$
\begin{equation}
B\equiv v \left[
\begin{array}{cccc}
0 & 0 & 0 & -b/v \\ 0 & 0 & 2b/v & 0 \\ 0 & v/2b & 0 & 0
\\ -v/b & 0 & 0 & 0
\end{array}
\right]
\end{equation}
The choice of $ b = v^{-8} = 31.152$, gives
\begin{equation} B\equiv v
\left[
\begin{array}{cccc}
0 & 0 & 0 & -v^{-9} \\ 0 & 0 & 2v^{-9} & 0 \\ 0 & v^{9}/2 & 0 & 0
\\ -v^{9} & 0 & 0 & 0
\end{array}
\right]
\end{equation}
and corresponds to the mass relation $ m_{b}
=\frac{m_{t}}{b}\left[ 1-\frac{vy}{\sqrt{2}}\right] =4.407...GeV $
for $m_t = 174.3 GeV$.

\section{Commutator plus anti-commutator symmetry algebras:}

The anti-commuting matrices
\begin{equation}
m\equiv \left[
\begin{array}{cc}
1 & 0 \\ 0 & -1
\end{array}
\right] ,p\equiv \left[
\begin{array}{cc}
0 & -a/v \\ v/a & 0
\end{array}
\right] ,q\equiv \left[
\begin{array}{cc}
0 & a/v \\ v/a & 0
\end{array}
\right]
\end{equation}
satisfy $[m,p]=-2q,[m,q]=-2p,[p,q]=-2m$. \ Similarly, $m$ and
\begin{equation}
r\equiv \left[
\begin{array}{cc}
0 & -v/2a \\ 2a/v & 0
\end{array}
\right] ,s\equiv \left[
\begin{array}{cc}
0 & v/2a \\ 2a/v & 0
\end{array}
\right]
\end{equation}
are anti-commuting and satisfy $%
[m,r]=-2s,[m,s]=-2r,[r,s]=-2m$. Note $m^2=q^2=s^2=1$,
$p^2=r^2=-1$, and that $a$ is arbitrary.  Consequently, if one
does not distinguish the $(+)$ versus SM indices, respectively of
the rows and columns, the tWb-transformation matrices have some
simple properties:

The anticommuting 4x4 matrices
\begin{equation}
M\equiv v\left[
\begin{array}{cc}
m & 0 \\ 0 & -m
\end{array}
\right] ,P\equiv v\left[
\begin{array}{cc}
-p & 0 \\ 0 & r
\end{array}
\right] ,Q\equiv v\left[
\begin{array}{cc}
q & 0 \\ 0 & s
\end{array}
\right]
\end{equation}
satisfy the closed algebra
$[\overline{M},\overline{P}]=2\overline{Q},[\overline{M},\overline{Q}]
=2\overline{P},[\overline{P},\overline{Q}]=2\overline{M}$. The bar
denotes removal of the overall ``$v$" factor, $M= v \overline{M},
...$.  Note that $Q$ is not a tWb-transformation.

Including the B matrix with $b$ arbitrary, the algebra closes with
3 additional matrices
\begin{equation}
\overline{B}\equiv \left[
\begin{array}{cc}
0 & d \\ f & 0
\end{array}
\right] ,\overline{C}\equiv \left[
\begin{array}{cc}
0 & e \\ g & 0
\end{array}
\right]
\end{equation}

\begin{equation}
\overline{G}\equiv \left[
\begin{array}{cc}
0 & h \\ k & 0
\end{array}
\right] ,\overline{H}\equiv \left[
\begin{array}{cc}
0 & j \\ l & 0
\end{array}
\right]
\end{equation}
where
\begin{equation}
d\equiv \left[
\begin{array}{cc}
0 & -b/v \\ 2b/v & 0
\end{array}
\right] ,e\equiv \left[
\begin{array}{cc}
0 & b/v \\ 2b/v & 0
\end{array}
\right] ,f\equiv \left[
\begin{array}{cc}
0 & v/2b \\ -v/b & 0
\end{array}
\right] ,g\equiv \left[
\begin{array}{cc}
0 & v/2b \\ v/b & 0
\end{array}
\right]
\end{equation}

\begin{equation}
h\equiv \left[
\begin{array}{cc}
-2ab/v^{2} & 0 \\ 0 & b/a
\end{array}
\right] ,j\equiv \left[
\begin{array}{cc}
2ab/v^{2} & 0 \\ 0 & b/a
\end{array}
\right] ,
 \newline
k\equiv \left[
\begin{array}{cc}
1/2v^{2}ab & 0 \\ 0 & -a/b
\end{array}
\right] ,l\equiv \left[
\begin{array}{cc}
1/2v^{2}ab & 0 \\ 0 & a/b
\end{array}
\right]
\end{equation}
The squares of the $2$x$2$ matrices (24-25) do depend on $a$, $b$,
and $v$.

The associated closed algebra is: $[\overline{M},\overline{B}]=0,\{\overline{%
M},\overline{B}\}=-2\overline{C};[\overline{B},\overline{C}]=0,\{\overline{B}%
,\overline{C}\}=-2\overline{M};$ \newline
$[\overline{M},\overline{C}]=0,\{\overline{M},%
\overline{C}\}=-2\overline{B};$ and
$[\overline{P},\overline{B}]=2\overline{H}
,\{\overline{P},\overline{B}\}=0;[%
\overline{H},\overline{P}]=2\overline{B},\{\overline{H},\overline{P}\}=0;$
\newline
$[%
\overline{H},\overline{B}]=2\overline{P},\{\overline{H},\overline{B}\}=0$
. Similarly,  $[\overline{P},\overline{C}]=0,\{\overline{P},\overline{%
C}\}=-2\overline{G};[\overline{M},\overline{H}]=-2\overline{G},$
\newline
$ \{\overline{M}%
,\overline{H}\}=0;[\overline{H},\overline{C}]=0,\{\overline{H},\overline{C}%
\}=2\overline{Q};$ and
$[\overline{M},\overline{G}]=-2\overline{H},\{%
\overline{M},\overline{G}\}=0;[\overline{P},\overline{G}]=0,$
\newline
$\{\overline{P},\overline{G}\}=2\overline{C};[\overline{G},\overline{B}]=-2%
\overline{Q},\{\overline{G},\overline{B}\}=0;$ and  $[\overline{G},\overline{C%
}]=0,\{\overline{G},\overline{C}\}=-2\overline{P};$ \newline $[\overline{G},\overline{H}%
]=2\overline{M},\{\overline{G},\overline{H}\}=0.$ The part involving  $%
\overline{Q}$ \  is
$[\overline{G},\overline{Q}]=2\overline{B},\{\overline{G},\overline{Q}\}=0;[%
\overline{B},\overline{Q}]=2\overline{G},$ \newline
 $ \{\overline{B},\overline{Q}\}=0;
 [\overline{C},\overline{Q}]=0,
 \{\overline{C},\overline{Q}\}=-2\overline{H};$ $%
[\overline{H},\overline{Q}]=0,\{\overline{H},\overline{Q}\}=
2\overline{C}$.

\bigskip

\ This has generated an additional tWb-transformation $G\equiv v\overline{G}$%
; but $C\equiv v\overline{C}$ and $H\equiv v\overline{H}$ are not
tWb-transformations. \

\bigskip

Up to the insertion of an overall $\iota =\sqrt{-1}$, each of
these 4x4 barred matrices is a resolution of unity, i.e.
$\overline{P}^{-1}=-\overline{P}$,
$\overline{G}^{-1}=-\overline{G}$, but
$\overline{Q}^{-1}=\overline{Q}$,
$\overline{B}^{-1}=\overline{B},...$ .

\section{Discussion:}

\indent {\bf (1)  $\Lambda_{+}$ mass scale:}

A fundamental question [2] raised by the existence of the
intensity-ratio equivalence of two couplings for the $t
\rightarrow W^{+} b $ mode is ``What is the origin of the
$\Lambda_{+} =E_W /2 \sim 53 GeV$ mass scale?" The present paper
shows that $M$ is but one of three logically-successive
tWb-transformations which are constrained by the helicity
amplitude ratio-relations (i) and (ii). Thereby, the type (iii)
ratio-relation fixes $\Lambda_{+} = E_W/2$ and the overall scale
of the tWb-transformation matrix $M$. The amplitude condition
(iv), $A_{+} (0,-1/2) = a A_{SM} (-1,-1/2)$ with $a= 1 + O(v \neq
y \sqrt{2}, x)$, and the amplitude condition (v), $A_{+} (0,-1/2)
= - b A_{SM} (1-1/2)$ with $ b = v^{-8} $, determine respectively
the scale of the tWb-transformation matrices $P$ and $B$ and
characterize the values of $m_W/m_t$ and $m_b/m_t$. The overall
scale can be set by choosing either $ m_t $ or $ {m_W}$. From an
empirical ``bottom-up" perspective of further ``unification", $
m_W$ is more appropriate to use to set the scale since its value
is fixed in the SM by the vacuum expectation value of the Higgs
field, $\phi$.

From the $M$ transformation, the $\Lambda_{+}$ mass scale is fixed
as $E_W/2$. From all three tWb-transformations, the numerical
value of $\Lambda_{+} $ is determined by that of $v_{EW}$.
Consequently like the value of the W-boson mass, the
dimensional-analysis required $\Lambda_{+}$ scale of the $t_R
\rightarrow b_L$ chiral weak transition-moment is not {\it
in-itself} a ``new mass scale" such as arising from a SUSY or a
technicolor generalization of the SM, but instead the value of
$\Lambda_{+}$ is another manifestation of the SM's electroweak
scale.

{\bf (2)  Comparison with an Amplitude Equivalence-Theorem:}

Given the continued successes of predictions based on the
couplings and symmetries built into the SM, and given the present
rather slow pace of new experimental information, we appreciate
the fact that for many readers it can be difficult to remind
oneself that directly from experiment we still really do not know
much about the properties of the on-shell t-quark [1]. Because of
possible form factor effects and possible unknown thresholds due
to new particles, in a theory/model-independent manner one cannot
reliably determine the indirect constraints on on-shell t-quark
couplings from off-shell contributions from t-quark contributions
in higher-order loops in electroweak precision tests.  Various
assumptions of quark universality are still routinely made in the
theoretical literature to conjecture on-shell t-quark properties
from theoretical patterns found for the several-order-of-magnitude
less massive quarks, in spite of the closeness of the value of
$m_t$ to $v_{EW}$ and of the now significantly greater
numerical-precision ($3\%$) of the $m_t$ measurement than that of
any of the other quark masses.

This present ``not-knowing" status quo for the on-shell t-quark is
very different from that in 1940-1952 in regard to the then
rapidly changing and developing experimental status quo in the
case of the weak and strong interactions, which was concurrent
with the series of striking empirical-theoretical successes with
QED ( QED is often viewed as the prototypical earlier analogue of
the present SM ). Nevertheless, despite these differences in the
experimental situation, we think it is instructive to compare this
present intensity-ratio equivalence for two distinct t-quark decay
couplings with a somewhat analogous ``amplitude equivalence
theorem (ET)" which was discovered and quite intensely studied,
circa 1940-1952, for the pseudoscalar and pseudovector
interactions, Dyson (1948) [9].

Although not as influential as Fermi's 1934 paper and the
Gamow-Teller 1936 paper concerning the Lorentz structure of the
weak interactions, this ET has had a long and significant impact
in high energy experimental and theoretical physics. The early
history of the ET , e.g. [9] and [10], can be traced from
Schweber's book (1962) [11] and from the entire final chapter of
Schweber, Bethe, and de Hoffmann, (1955) [12]. This ET stimulated
in part, the development of effective Lagrangian methods [13] and
work by G. t'Hooft and M. Veltman [14].  ( Later ET literature
cites an t'Hooft-Veltman preprint; the corresponding published
tHV-paper does not cite, e.g. [12], but refers to their tHV-paper
as being based on unpublished preprints. ) The ET continues to be
of theoretical interest, see for instance [15].

There are similarities and qualitative differences between this
intensity-ratio equivalence of two t-quark couplings and the
pseudoscalar-pseudovector ET. They are alike in that both relate
simple Lorentz structures, involve helicities...though more
intricately in the present case, and involve a renormalizable
coupling. Major differences include (I) the ET case is much better
understood after over 60 years of research, whereas this paper is
only a beginning towards understanding the symmetries and possible
deeper physics implications/structure of this coupling-equivalence
in t-quark decay, (II) the SM is now known to well explain most of
the weak interaction systems (nucleon, nucleus, strange particle
weak decays) first studied by the ET stimulating experiments,
whereas precision experiments have only begun on t-quark decay,
and (III) the tWb transformations involve couplings of a
fundamental renormalizable local quantum field theory, the SM, and
fundamental mass ratios, whereas, instead, the ET case involved
hadronic couplings.

{\bf (3) Possible Implications of These Symmetries:}

The presence in nature of an additional $t_R \rightarrow b_L$
weak-moment coupling would be but a further extension of the
observed $(V-A)$ chiral asymmetry, i.e. a $(V+A) + (f_M - f_E)$
coupling versus a $(V+A)$ one would also lead to similar five
types of analytical relations and a possible large $(f_M - f_E)$
chiral weak moment [8]. However, the additional $(f_M + f_E)$
coupling does violate the conventional gauge invariance
transformations of the SM and in the past, in electroweak studies
such anomalous couplings have been best considered as ``induced"
or ``effective". Nevertheless, in special ``new physics"
circumstances such a simple charge-changing tensorial coupling as
(2) might turn out to be a promising route to deeper
understandings. Observation of a charge-changing ``tensorial
coupling" could prove to be a fundamental step forward from
gravitation viewpoints.

An important formal, field-theoretic question is whether the new
symmetries associated with the $(V-A) + (f_M + f_E)$ chiral
structure such as the symmetry algebras of Sec. 3 are sufficient
to overcome the known difficulties [16] in constructing a
renormalizable, unitary quantum field theory involving second
class currents [17] ? The $f_E$ component is second class. What
might be of deeper significance in the context of CP violation, is
that $f_E$ has a distinctively different reality structure, and
time-reversal invariance property versus the first class $ V, A,
f_M$ components [18].

During 1970-1980 and later for tau-decay processes where $m_{\tau}
\gg m_{\nu}$, various refinements in the classification of
possible second class weak-interaction currents were suggested.
Specific mechanisms/models were proposed for introducing such
currents into the standard theory of the weak interactions [19].
Phenomenologically, the tWb-transformations can be viewed as
another such attempt to generalize beyond the weak interaction
couplings as embodied in the SM.  If the observed $t \rightarrow
W^{+} b $ decay mode is found empirically to be well described by
(2), this would then support a bolder working premise that in the
on-shell limit of any treatment of t-quark decay and of $m_t$ and
$m_b$, the underlying symmetries of these tWb-transformations are
basic to relating the associated mass scales, much as are Lorentz
invariance and the gauge-symmetries of $ SU(2)$X$U(1)$ dynamics in
performing calculations in the SM.

{\bf (4)  Experimental Tests/Measurements:}

For $t \rightarrow W^{+} b$ decay channel there is a sizable $v^2$
factor difference of partial widths: the SM's $\Gamma_{SM}= 1.55
GeV$, versus $\Gamma_+ = 0.66 GeV$ and a longer-lived (+) t-quark
if this mode is dominant.  Complimentary to this on-shell process
are the s-channel and t-channel off-shell processes in
single-top-production at the Tevatron and LHC.  If the SM
description of t-quark phenomena is completely correct,
measurement of CKM factor $|V_{tb}|^2$ will be possible via the
s-channel process $ q q^{\prime} \rightarrow t \bar{b} $ in
single-top production [20]. At present, from $162$ $pb^{-1}$ of
data, the CDF bound is $ 13.6 $pb versus the theoretical $ 0.88
\pm 0.11$pb (SM) prediction.  From this, there is the estimate
that the single-top-production process will be observed at the
Tevatron when $1-2$ $fb^{-1}$ of data has been accumulated [1].
However, as is emphasized elsewhere in this paper, without
assumptions concerning possible new particle thresholds and the
off-shell behaviors of form factors, one can not reliably predict
from $t \rightarrow W^{+} b$ decay couplings what will occur in
higher order, off-shell weak-interaction precision tests nor in
crossed channel reactions. In the literature, direct experimental
tests for other than $(V-A)$ couplings in single-top-production
have been proposed for both the s-channel and the t-channel
(gluon-W fusion)[21].  In contrast to past searches for anomalous
weak-interaction couplings for the other quarks and for the
leptons, because the top mass is so large, the kinematics for all
three of these processes (decay channel, single-top-production
s-channel and t-channel) is excellent for direct searches for
anomalous couplings.  However, concurrently versus applications in
light-quark and leptonic reactions, large and uncertain
dispersion-theoretic extrapolations will be required in any
attempt to relate measurements between these different t-quark
reactions.

In on-going [1] and forth-coming [6,7] $t \rightarrow W^{+} b$
experiments, important information about the relationship of the
tWb-transformation symmetry patterns of this paper to the observed
t-quark decay channel will come from:
\newline
(a) Tests for the existence of anomalous couplings and if found,
for their structure and symmetries, such as from the on-going
CDF/D0 analyses which investigate the W-boson helicity [1,4,22].
\newline
(b) Measurement of the sign of the $\eta_L \equiv \frac 1\Gamma
|A(-1,-\frac 12)||A(0,- \frac 12)|\cos \beta _L = \pm 0.46(SM/+) $
helicity parameter [4] so as to determine the sign of $cos \beta_L
$ where $ \beta_L = \phi _{-1}^L- \phi _0^L $ is the relative
phase of the two $\lambda_b = - \frac{1}{2} $ amplitudes,
$A(\lambda _{W^{+}},\lambda _b)=|A|\exp (i\phi _{\lambda
_{W^{+}}}^{L,R})$.  For the exclusion of the coupling of (2)
versus the SM's $(V-A)$ coupling, this would be the definitive
near-term measurement concerning properties of the on-shell
t-quark.
\newline
(c) Measurement, or an empirical bound, for the closely associated
\newline $ {\eta_L}^{'} \equiv \frac 1\Gamma |A(-1,-\frac
12)||A(0,- \frac 12)|\sin \beta _L $ helicity parameter.  This
would provide useful complementary information, since in the
absence of $T_{FS}$-violation, ${\eta_L}^{'} =0$ [4].

Since the helicity amplitude relations discussed in Sec. 2 involve
the b-quark helicities, c.f. differing signs in $\lambda_b = 1/2$
column of Table 1, there are also independent phase tests which
require
\newline
(d) Measurements of helicity parameters [5] using
$\Lambda_b$-polarimetry in stage-two spin-correlation functions.
It is noteworthy that the $\Lambda_b$ baryon has been observed by
CDF at the Tevatron [23].

For the case of spin-correlations in SM $ t \bar{t}$ pair
production and decay, several groups have investigated various
higher-order corrections [24].

{\bf Acknowledgments: }

We thank experimental and theoretical physicists for helpful and
intellectually stimulating discussions, especially ones at ``Vth
Recontres du Vietnam 2004" and at ICHEP04 in Beijing. This work
was partially supported by U.S. Dept. of Energy Contract No. DE-FG
02-86ER40291.

{\bf Appendix: The $O(v \neq y \sqrt{2}, x)$ corrections in $a$}

In this appendix is listed the form of the $O(v \neq y \sqrt{2},
x)$ corrections in $a$ as required by Lorentz invariance:

For $a=1+\varepsilon (x,y)$, the (iv) relation is $v=(1+\varepsilon )y \sqrt{2}%
m_{t}/(E_{W}+q)$ whereas from relativistic kinematics $\ \
v=q/E_{W}=[(1-y^{2}-x^{2})^{2}-4y^{2}x^{2}]^{1/2}/[1+y^{2}-x^{2}]$
. \ By equating these expressions and expanding in $x$, one
obtains $ \varepsilon =R+x^{2}S $ where
\begin{eqnarray*}
R &=&\frac{1-4y^{2}-3y^{4}-2y^{6}}{4y^{2}(1+y^{2})^{2}} \\ S
&=&\frac{-1-4y^{2}+y^{4}}{2y^{2}(1+y^{2})^{3}}
\end{eqnarray*}
and $  v= y \sqrt{2} \left[
1+R+x^{2}(S+\frac{1+R}{1-y^{2}})+O(x^{4})\right] $.  From the
latter equation, $R=(v-y\sqrt{2})/y\sqrt{2}+O(x^{2}).$

For a massless b-quark ( $x=0$ ) and $a=1$ , the (iv) relation is
equivalent
to the $\frac{m_{W}}{m_{t}}$ mass relation \  $y^{3}\sqrt{2}$ $+y^{2}+y\sqrt{%
2}-1=0$, and by relativistic kinematics to the W-boson velocity condition $%
v^{3}+v^{2}+2v-2=0$ and the simple formula $v=y\sqrt{2}$.

\newpage

\begin{center}
{\bf Table Captions}
\end{center}

Table 1:  Numerical values of the helicity amplitudes for the
standard model $(V-A)$ coupling and for the (+) coupling of Eq.(2)
which has a $(V-A) + (f_M + f_E)$ Lorentz-structure. The latter
coupling consists of an additional $t_R \rightarrow b_L$ chiral
weak transition-moment of relative strength $\Lambda_{+} \sim 53
GeV$ so as to produce a relative-sign change in the $\lambda_b= -
\frac{1}{2}$ amplitudes. The values are listed first in $ g_L =
g_{f_M + f_E} = 1 $ units, and second as $ A_{new} = A_{g_L = 1} /
\surd \Gamma $. Table entries are for $m_t=175GeV, \; m_W =
80.35GeV, \; m_b = 4.5GeV$.

\end{document}